# Geomagnetic Earthquake Precursors Improvement Formulation on the basis of SKO (Skopje) and PAG (Intermagnet) Geomagnetic Data

**Strachimir Chterev Mavrodiev** [a,*], **Lazo Pekevski** [b]

[a] Institute for Nuclear Research and Nuclear Energy, Bulgarian Academy of Sciences, Bul. Tzarigradsko shose 72, Sofia 1784, Bulgaria.
[b] Seismological Observatory, Faculty of Natural Sciences and Mathematics, Sts. Cyril and Methodius University, Skopje, Macedonia.

**ABSTRACT**

In this paper we present that the simple analysis of the local geomagnetic field behaviour can be a reliable imminent precursor for regional increasing seismicity. In the first step the problem was investigated using the one-component Dubna fluxgate magnetometer. The result of 2001-2004 Sofia monitoring confirmed many older papers for connection between Earth tide (Sun-Moon tides as earthquake's trigger) and jump (Geomagnetic quake) of daily averaged of one minute stndart deviation of the geomagnetic field. The second step (2004-present), which included a three-component Danish fluxgate magnetometer, worked in Skopje Seismological observatory confirmed the first step result. The analysis of INTERMAGNET data stations around which was happened stronger earthquakes also confirmed our result. The distribution of time difference between the times of such earthquakes and local daily averaged tide's vector movement for impending tide's extremum confirms our estimate that the increasing seismicity is realized in time window about +/- 2.7 days.

The Complex program for researching the possibility for "when, where and how" earthquake's prediction is proposed as well as the short description of FORTRAN codes for analysis of earthquakes, geomagnetic and tide data, their correlations and visualization.

## 1. Introduction.

The hypothesis for possible correlations between the earthquakes, the variations of magnetic fields, Earth's horizontal and vertical currents in the atmosphere, was born when in early 1988, the historical data on the Black Sea was systemized [Mavrodiev, 1996].

The achievement in the Earth surface tidal potential modeling, with the ocean and atmosphere tidal influences being included, makes an essential part of the research. In this sense, the comparison of the Earth tides analysis codes (Venedikov et al., 2003; Milbert, 2011) is very useful. The possible tidal triggering of earthquakes has been investigated for a long period of time (Knopoff, 1964; Tamrazyan, 1967, 1968; Ryabl at al., 1968; Shlien, 1972; Molher, 1980; Sounau et al., 1982; Burton, 1986; Shirley, 1988; Bragin, 1999).

The last-years laboratory results in modeling the earthquake processes in increasing stress conditions support, at least qualitatively, the quantum mechanic phase shift explanation of the mechanism of generation of electromagnetic effects before and during earthquakes (Freund et al., 2002; St-Laurent et al., 2006; Vallianatos et al., 2003).

Some progress in establishing the geomagnetic filed and Eart tides variations as imminent precursors for increasing regional seismicity was presented in several papers (Mavrodiev, Thanassoulas, 2001; Mavrodiev, 2002a, b; 2003a, b, c; Mavrodiev, 2004; Mavrodiev, Pekevski, 2008; Mavrodiev, Pekevski, Jimseladze, 2008).

The earthquake-related part of the models has to be infinitely repeated in the "theory–experiment–theory" process, using nonlinear inverse problem methods in looking for correlations between the different fields in dynamically changing space and time scales. Each approximate model supported by some experimental evidence should be included in the analysis (Varotsos et al., 2006; Thanassoulas et al., 2001a, b; Eftaxias at al., 2001, 2002; Duma, 2006). The adequate physical understanding of the correlations among electromagnetic precursors, tidal extremes and a impendant earthquake is related to the progress of an adequate Earth magnetism theory and electrical currenets distribution, as well as to the quantum mechanical understanding of the processes in the earthquake source volume before and during the earthquake.

Simultaneous analysis of more accurate space and time measuring sets for the Earth crust condition parameters, including the monitoring data of the electromagnetic field under and over the Earth surface, as well as the temperature distribution and other possible precursors, would be the basis of nonlinear inverse problem methods. It could be promising for studying and solving the "when, where and how" earthquake prediction problem.

## 2. The discovery of geomagnetic quake as reliable precursor for increasing of regional seismicity

In December 1989, a continuous measurement of a projection of the Earth's magnetic field (F) with a magnetometer (know-how of JINR, Dubna, Boris Vasiliev) with absolute precision less than 1 nano-Tesla at a sampling rate of 2.5 samples per second was started. The minute's mean value of *F*, its error mean value, the minute's standard deviation *SDF*, and its error were calculated, i.e., every 24 hours, 1440 quartets of data were recorded.
Minute standard deviation of *F* is defined as:

$$SDF_m = (1/N_m) \sum_{i=1}^{N_m} (F_i + F_{mean})^2 \tag{1}$$

where $F_{mean} = (1/N_m) \sum_{i=1}^{N_m} F_{mean}$ (2)

and $N_m$ is the number of samples per minute (150).

The connection between variations of local geomagnetic field and the Earth currents was established in INRNE, BAS, Sofia, 2001 seminar [1]. The statistic of earthquakes that occurred in the region (1989- 2001), confirmed the Tamrazyan notes [2, 3], that the extrems of tides are the earthquake's trigger. The Venedikov's code [4, 5] for calculating the regional tide force was used [Mavrodiev, 2004]

The signal for immient increasing regional seismic activity (incoming earthquakes) is the "geomagnetic quake" (Gq), which is defined as a jump (positive derivative) of daily averaged *SDF*. Such approach permits to compare by numbers the daily behaviour of the geomagnetic field with those in other days.

Among the earthquakes occurred on the territory under consideration in certain time period, the "predicted" one is the earthquake with magnitude $M_P$ and epicentral distance $Dist_P$ which is identified by the maximum value of the function:

$$S_{ChM} = 10^{1.5M+4.8} / (D + Depth_P + Dist_P)^2 \quad [energy/km^2] \tag{3}$$

where D=40 km is fit a parameter.

The physical meaning of the function S<sub>ChtM</sub> is a surface density of earthquake's energy in the point of measurement. It is important to stress out that the first consideration of the earthquake magnitude *M* and epicentral distance dependence was obtained using nonlinear inverse problem methods. Obviously, the close distance strong earthquake (with relatively high value of S<sub>ChtM</sub>) will bear more electropotential variations, which will generate more power geomagnetic quake.

It is very important to note that in the time scale of one minute, the correlation between the time period of increasing regional seismic activity (incoming earthquake), and tide extrema, recognized of predicted earthquake was established using the Alexandrov's code REGN for solving the over determined nonlinear systems [6, 10]. The very big worthiness of Alexandrov's theory and code is possibility to choose between two functions, which describe the experimental data with the same hi-squared, the better one.

## 3. The existence of geomagnetic quake in a case of geomagnetic vector monitoring

At the SKO (Seismological Observatory Observatory in Skopje) is installed triaxial fluxgate magnetometer FGE model (Danish Meteorological Institute), with year sensors drift less then a few nT. The magnetometer is used in variometer mode with resolution 0.2 nT, and sampling by 10 samples per sec.

During routine geomagnetic data processing at SKO, *MinuteSig,* was calculated as mean minute value of 10 samples per second recorded geomagnetic variation data:

$$MinuteSig = \sqrt{\frac{SDH^2 + SDD^2 + SDD^2}{H^2 + D^2 + Z^2}} \qquad . \qquad (4)$$

So the daily signal *GmSig* was,

$$GmSig_{SKO} = \frac{1}{1440} \sum_{i=1}^{1440} MinuteSig_i \qquad . \qquad (5)$$

For PAG, the available data were already Intermagnet minute variations data (available by Intermagnet data service), so the *HourSig,* were calculated for one hour time interval,

$$HourSig = \sqrt{\frac{SDX^2 + SDY^2 + SDZ^2}{X^2 + Y^2 + Z^2}} \qquad (6)$$

and the daily signal *GmSig* was

$$GmSig_{PAG} = \frac{1}{24} \sum_{i=1}^{24} HourSig_i \qquad . \qquad (7)$$

The *precursor* signal in both cases was calculated by the formula:

$$\Pr ecursorSig = \frac{GmSig_{Today} - GmSig_{Yesterday}}{0.5 * (A_{Today} + A_{Yesterday})} \qquad , \qquad (8)$$

where $A_{Today}, A_{yesterday}$ stand for day by day geomagnetic activity middle A indices (provided by NOAA on http://www.swpc.noaa.gov/ftpdir/indices/DGD.txt ). So, the impact of Cosmos and Sun generated variations of global geomagnetic field is compensated and its influence on the calculations is avoided particulary.

Now, the "geomagnetic quake" Gq is a signal for increasing of the imminent regional seismic activity (incoming earthquakes) and has more precise definition than previously: a jump (positive derivative) of the *PreqursorSig* value. When this criterion is fulfilled, one can say that the geomagnetic quake has a place.

Such quake is unique precursor for incoming earthquakes and in the next minimum or maximum of the local Tidal gravitational extreme, daily averaged module R of vector movement, somewhere in the region under consideration such predicted earthquake/earthquakes will occur.

In the right part of the Fig.1., are presented, up to down minute averaged values of *H, D, Z*, and its standard deviation of *SDH, SDD, SDZ* correspodingly, as well as the variable *MinuteSig*, measured with 10 samples per second, and averaged on one minute.

In the right part of the Fig.2., are presented, up to down minute averaged values of *X, Y, Z*, and its standard deviation of *SDX, SDY, SDZ* correspodingly, as well as the variable *HourSig*, 1 minute samples and averaged on one hour.

In the left part of the both figures (1. and. 2.), down to up are presented as follow
- the values of *Middle A indices*, averaged daily middle latitude (Frederiksburg) geomagnetic activity indices.
- the next upper graph are presented the occurred earthquakes with index, which is the distance (<= 700 km) from the monitoring point.
- the next graph presents the daily sum of function *Schtm*, and
- the upper graphs presents the behaviour of minutes and daily averaged values of Earth's surface tidal movement, calculated with Dennis Milbert's code: http://home.comcast.net/~dmilbert/softs/solid.htm.

When the specific behavior of field and its standard deviation are occurred more than one time at different hours of the day, it means the signals from future earthquakes with different epicenters are recorded. This example is clearly presented on Fig.1 and Fig.2..

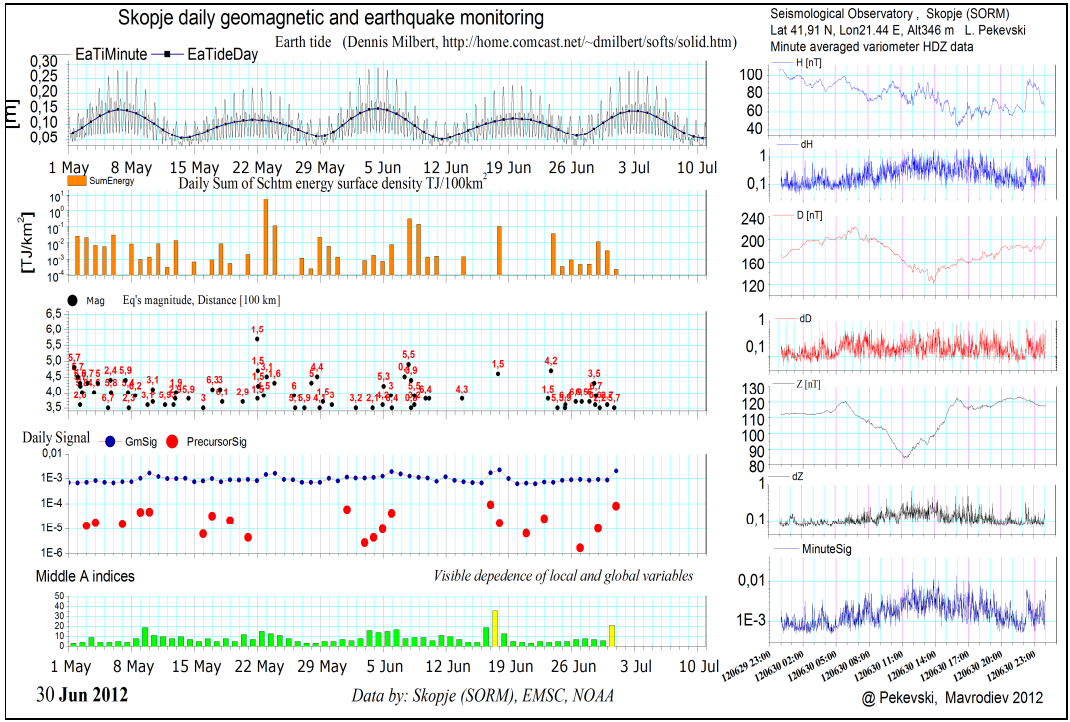

**Figure. 1**. Skopje daily geomagnetic and earthquake monitoring

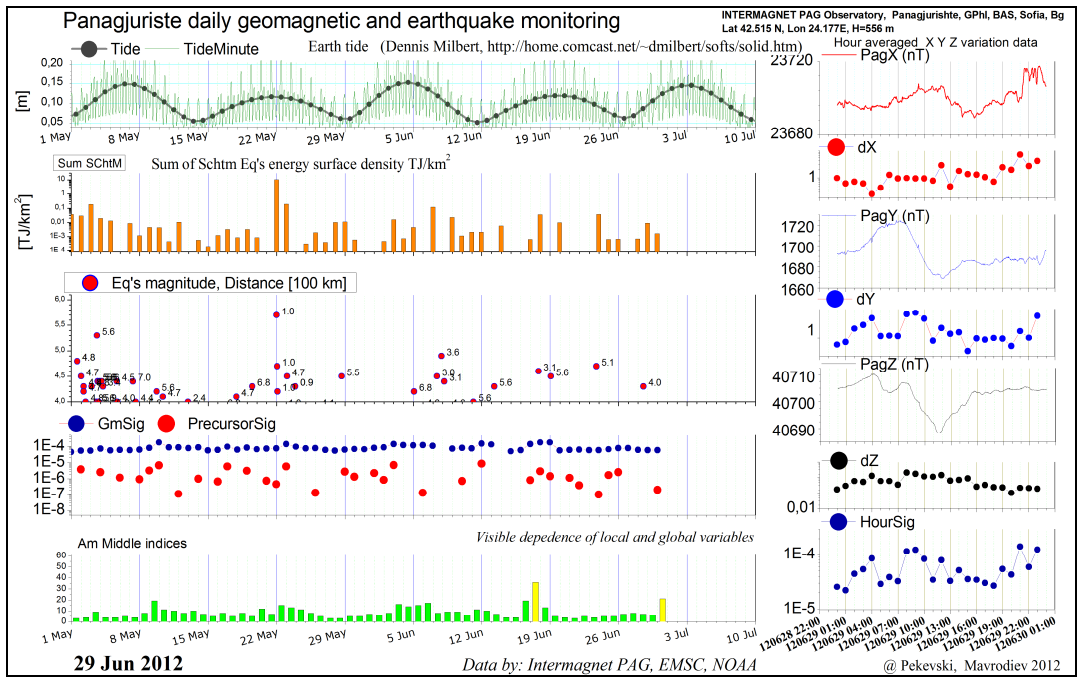

**Figure 2**. Panagjuriste daily geomagnetic and earthquake monitoring

The above Skopje and Panagjuriste figures illustrate how presented geomagnetic data processing methodology works for Mw 5.6, Pernik earthquake (May 22, 2012, 00:00 UTC).

## 4. Day-Difference analysis.

The role of the electromagnetic variations as earthquake's precursor can be explained in general by the hypothesis: the strain accumulation in the Earth Crust during the earthquake preparation causes medium's density change, within which a chemical phase ("dehidratation") shift and a corresponding electrical charges shift appears. The Earth tides extremum as earthquake's trigger could be based on the hypothesis of "convergence of tidal surface waves" in the region (territory with prominent tectonic activity as consequences of chemical phase shift) of impending seismic activity.

For every occurred earthquake was calculated "day-difference"; the smaller absolute time difference between the hypocental time and the daily times of pre and post tide's extremum time at that site on the Earth surface (the earthquake epicenter). This procedure was provided on all reported earthquakes in ISC catalogue (**http://www.isc.ac.uk/**) for the time period 1981-2011 and M≥3.5 and M≥4.0. The program for calculating of daily averaged module of vector movement $R_{mean}$ is based on Dennis Milbert TIDE programe (solid.for ), by which Tide data could be calculated only for the time period after 1981. The DailyTide time of the Tide extremes $R_{mean}$ were calculated by analogy of center mass calculation.

The statistic of day-differences for the earthquakes that occurred worldwide (1981- 2011) and the Gaussian fitted curve (Fig.3.), confirmed the Tamrazyan notes [2, 3] from 1960-th, that the extrems of tides play a role of earthquake's trigger.

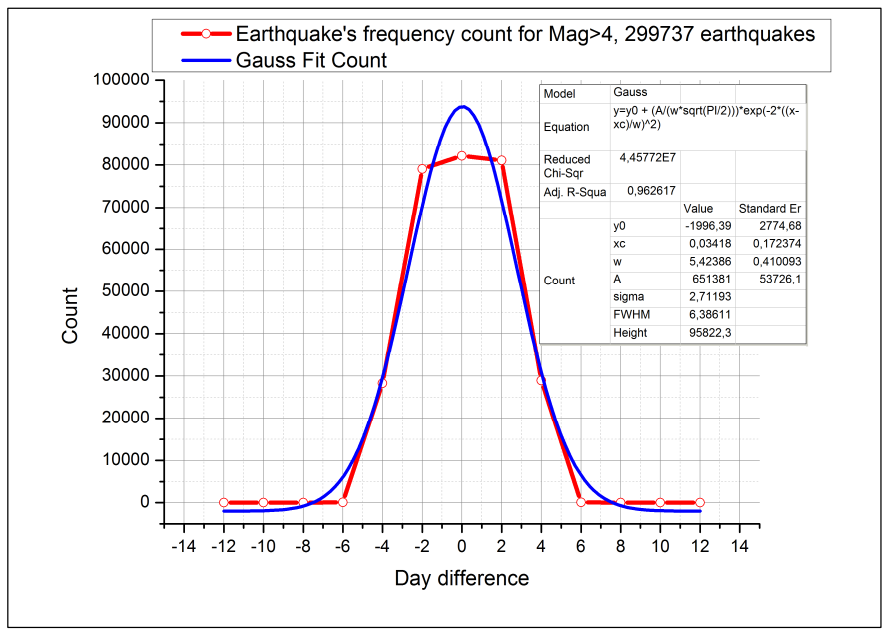

**Figure 3**. The distribution difference between the time of local tide extreme and the local hypocentral time of world occurred earthquakes (1981-2011, M>4.0) http://www.isc.ac.uk/ database.

## 5. SKO (Skopje) and PAG (Panagjurishte) day-difference statistic

Earthquakes data reported in ISC earthquake catalogue for the wider Balkan region in time period 2009-2011, were analyzed. The data included epicentral distances up to 400 km from Skopje (SKO) and Pangjurishte (PAG) sites and the lower earthquakes magnitude threshold was 3.5 (Fig.4).

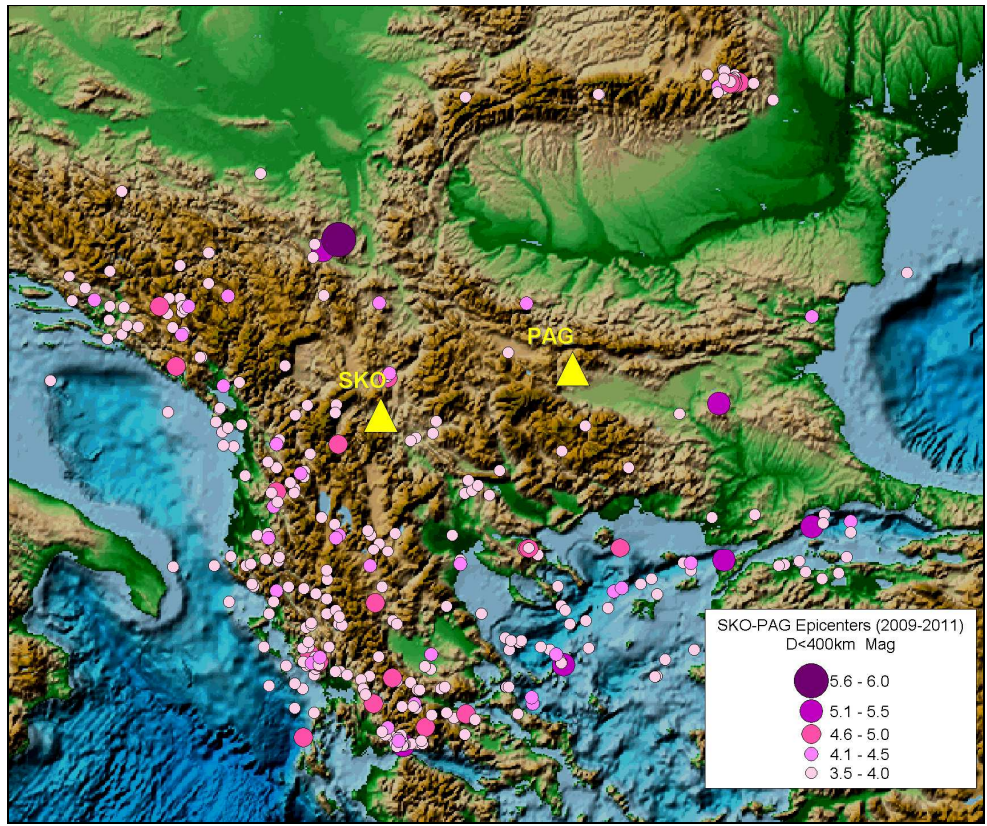

**Figure 4.** The epicentral map of earthquakes with M>3.5 for 2009-2011
epicentral distances D<400 km from SKO and PAG)

For the both sites on the basis of calculated day-difference data distributions for all occurred earthquakes (1981-2011, Fig.5.) as well for those with geomagnetic signals (2009-2011, Fig., 6.) are presented.

The facts that Gauss fits of all presented distributions (Fig.5-8) are similar and with good values of $\chi^2$ can be interpreted as confidence of our hypothesis for geomagnetic quake (Gq) existence as earthquakes regional immenent precursors.

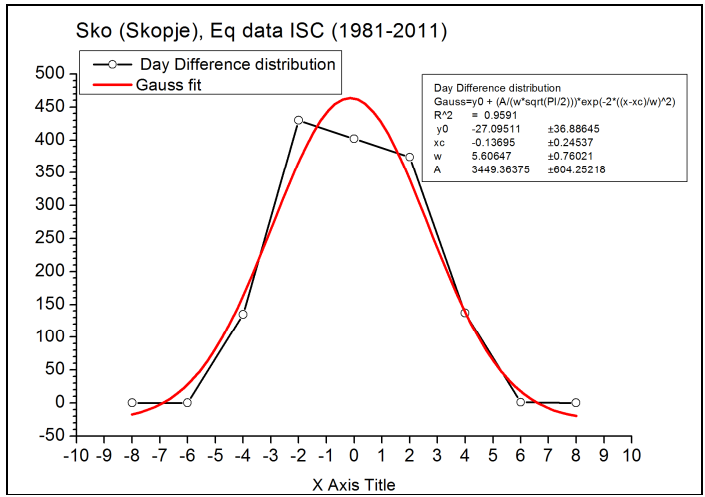

**Figure 5**. The day-difference data distributions for all occurred earthquakes (epicentral distances D<400 km from Seismological Observatory Skopje (SKO)

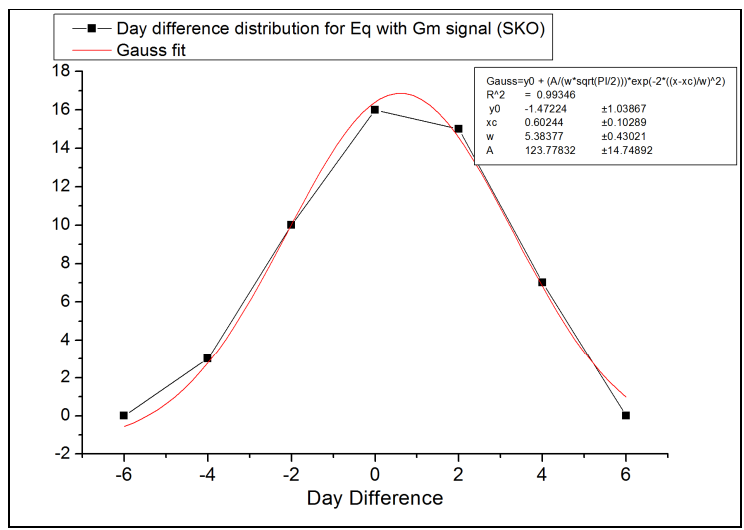

**Figure 6**. The day-difference data distributions occurred earthquakes with Gq signal (epicentral distances D<400 km from Seismological Observatory Skopje (SKO).

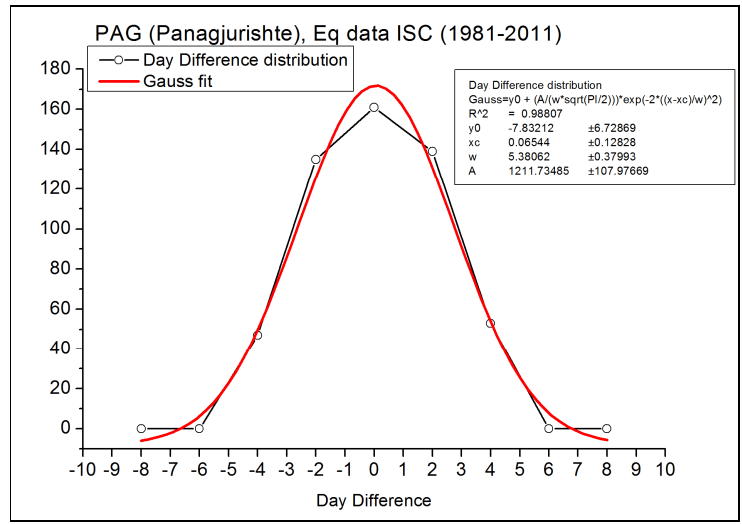

**Figure 7.** The day-difference data distributions for all occurred earthquakes Epicentral distances (D<400 km from Intermagnet Observatory Panagurishte (PAG))

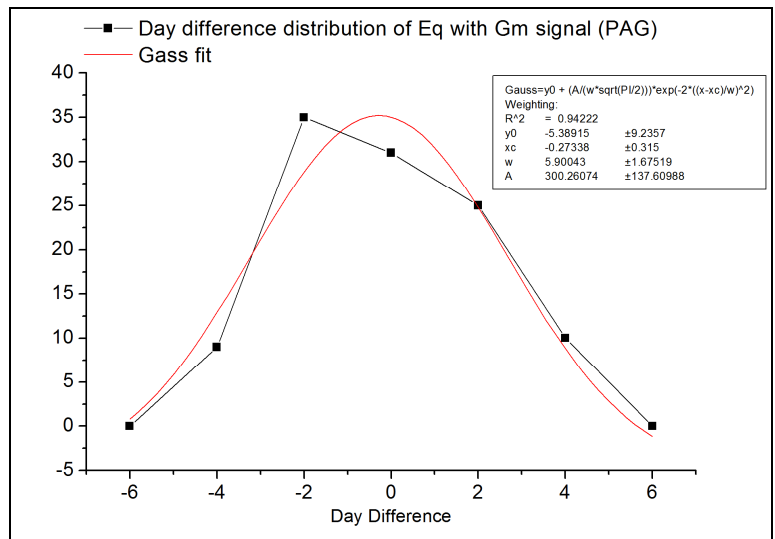

**Figure 8**. The day-difference data distributions for all earthquakes with Gq signal(epicentral distances(D<400 km from Intermagnet Observatory Panagurishte (PAG))

If one compares the number of all occurred earthquakes with the number of the earthquakes with Gm signals, it is evident the fact we have the same numbers for magnitudes M>4.9 in both cases, presented on Fig. 9 and Fig.10.

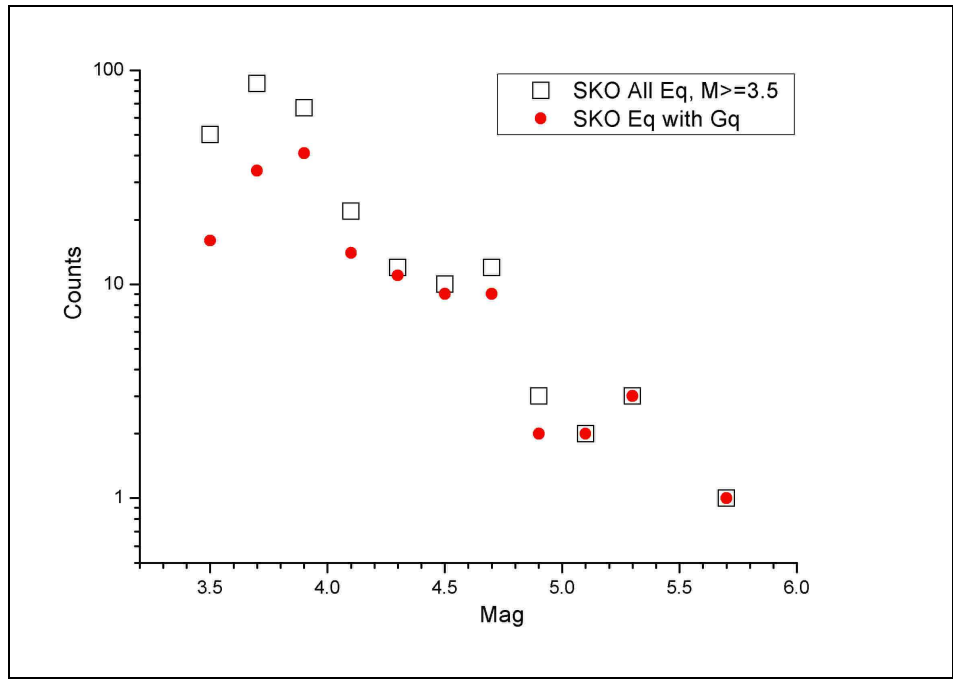

**Figure 9**. The comparison of the number of all occurred earthquakes and those with Gq signal
for Seismological Observatory Skopje (SKO).

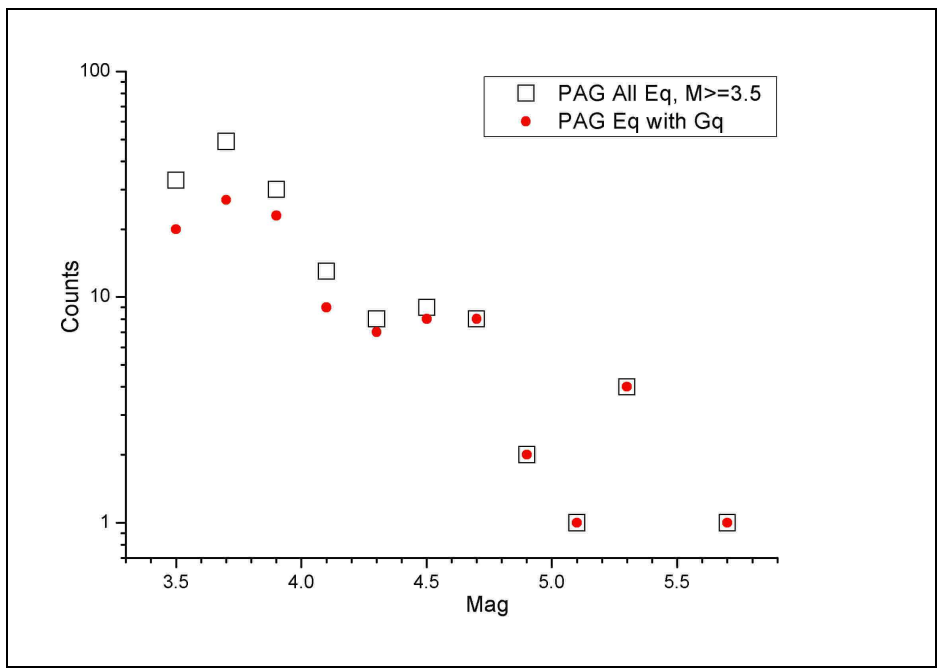

**Figure 10**. The comparison of the number of all occurred earthquakes and those with Gq signal for Panagurishte Intermagnet Observatory (PAG)
 I

# 6. Examples for apposteriory application of Gq approach for som big earthquakes

Such posteriori analysis on the basis of INTERMAGNET minute variations data for selected earthquakes also confirmed that the geomagnetic quake (Gq) is a reliable regional precursor for imminent seismic increasing seismic activity
.

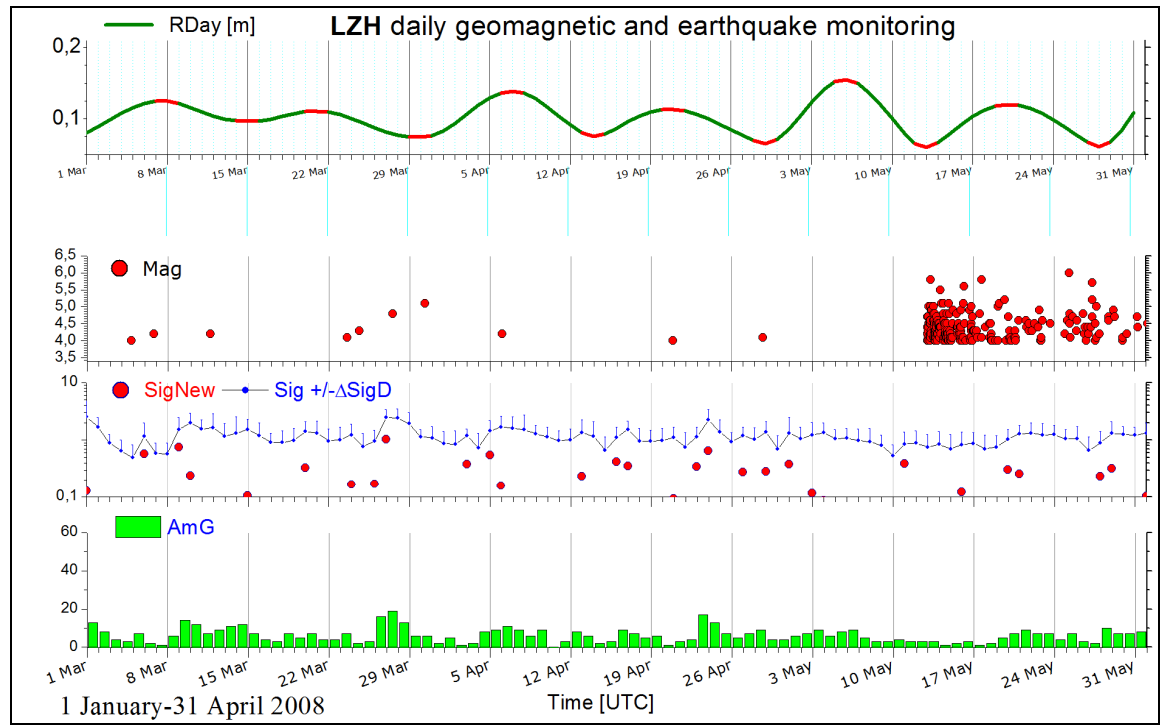

**Figure 11.** Daily geomagnetic monitoring and distribution of earthquakes for Intermagnet Observatory Lanzhou (LZH), China, and the earthquake of magnitude Mw 7.9 occurred on 12/05/2008 at 06:27:59.0 UTC in Wenchuan county, Easter Sichuan, China.

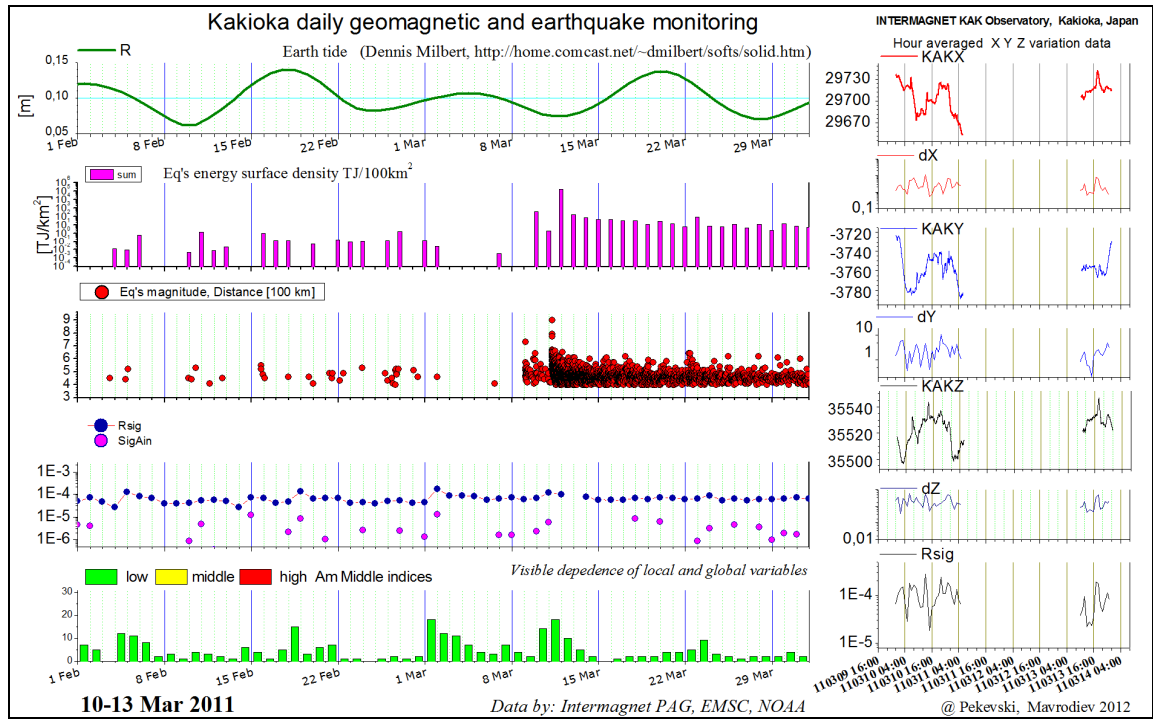

**Figure 12.** Daily geomagnetic monitoring and distribution of earthquakes for Intermagnet Observatory Kakioka (KAK), Japan, and the earthquake of magnitude Mw 9.0 occurred on 11/03/2011 at 05:46 UTC occurred off the Pacific coast of Tohoku, Honshu Island Japan.

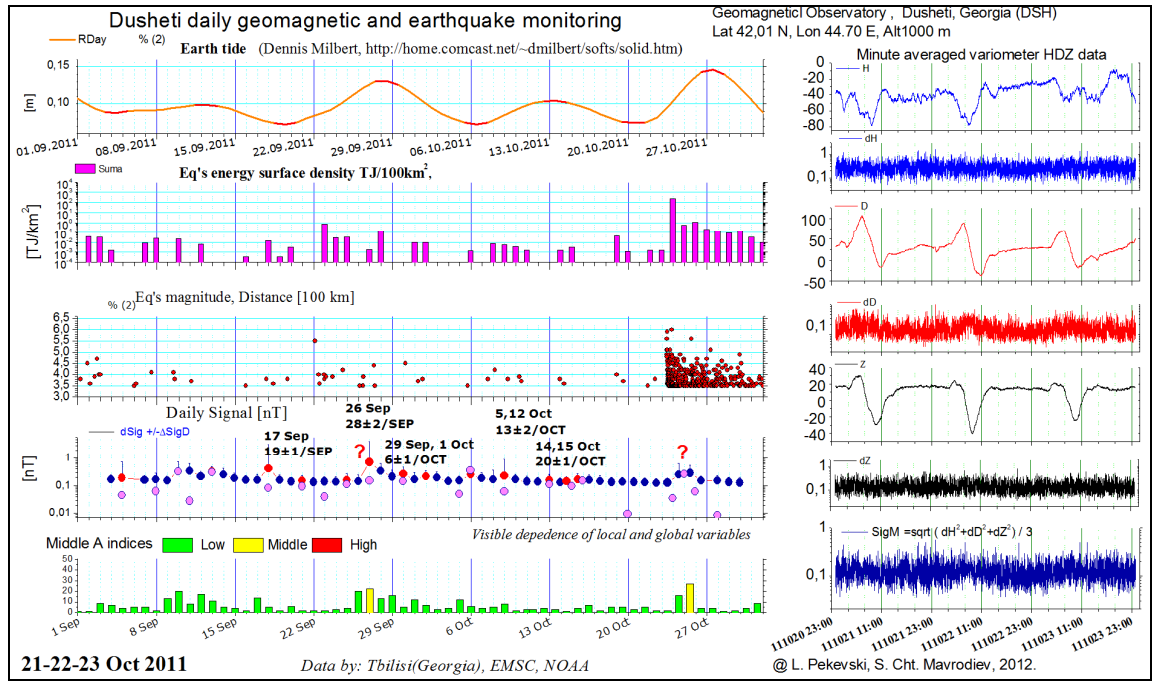

**Figure 13.** Daily geomagnetic monitoring and distribution of earthquakes for Geomagnetic Station in Dusheti, Georgia and the earthquake of magnitude Mw 7.2 occurred on 2011/10/23 at 10:41:22.0 UTC close to the city of Van in Eastern Turkey.

## 7. Proposal for complex Project for solving the "when, where and how" earthquake prediction

The purpose of the project is a development of long-term research cooperation through coordinated joint program for exchange of data, know-how and scientists. The partnership in experimental and theoretical aspects of geophysics will be focused on a creation of fundamentals of a Complex Program for investigation of the possibilities to forecast earthquake's time, hypocenter magnitude, and intensity using reliable precursors. For this aim, a monitoring of the following parameters is envisaged: geophysical and seismological variability over Region; water sources and their Radon, Helium, and other gases concentrations; crust temperature;electromagnetic field variations under, on and above Earth's surface;satellite monitoring of meteorological parameters, including earthquake clouds;electrical charge distributions and variations in ionospheric parameters;near-space monitoring, aimed to detect and exclude the external influences (i.e., those of Sun and Interplanetary medium variations, Cosmic rays) from the meaningful signal of the solid Earth;biological precursors.

    For many variables, well working global and regional monitoring exists (for example INTERMAGNET monitoring). For others (for example: i.e., monitoring of Earth's currents distribution) monitoring have to be created.

    The complex monitoring of the broad variety of parameters defines the output of the Program including: i) estimations of different time scales for more clearer understanding of the Earth's system natural variability, ii) risk assessment of the appearance of hazards for society events related to earthquakes, climatic changes, etc., and iii) people's response to an abrupt change in the monitored parameters.

The proposed regional network can be considered as a first step in creation of a wide interdisciplinary scientific consortium capable of formulation of a more adequate paradigm of climate variability and climatic change (distinguishing the differences between them), Earth seismic processes and the actual problem of their forecast. If the hypothesis for Georeactors (Rusov et al, 2006, 2009, 2010; Feoktistov, 1998; 238U; Teller, 1996; 232Th type) net, as possible reason for Climate change will be confirmed, the surprising new knowledge for Climate variations and its connection with Earth's seismicity can be obtained. Conformation of the hypothesis for existence of new type self-regulated nuclear reactors will enhance the physical understanding of different time scales of Climate - Climate and Climate - Weather transitions and Climate - Seismicity correlations.

**Conclusion**

The correlations between local geomagnetic quake and incoming earthquakes, which occur in the time window defined from the next minimum (+/- 1 day) or maximum (+/-2 days) of the Earth Tidal gravitational potential is tested statistically. The distribution of the time difference between predicted and occurred events is a Gauss one and is increase in the time.

This result can be interpreted like a possible first reliable approach for solving the "when" earthquakes prediction problem.

On the basis of electromagnetic monitoring under, on and over Earth surface is proposed research for solution of "when, where" earthquake prediction problem. Under the hypothesis the current has a big vertical component the data of two geomagnetic vector devices are enough for determination of the future epicenter. The three devices will permit to research the correlation between Earth surface distribution of precursor function Sig and the Magnitude of the incoming earthquake.

The complex Project for solving the "when, where and how" earthquake prediction problem is very shortly presented**.**

**Acknowledgement**

The results presented in this paper rely on the data collected at SKO (Skopje, Seismological Observatory) and PAG (Panagjurishte). We thank Geophysical Institute of the Bulgarian Academy of Science, for supporting PAG operation and INTERMAGNET for promoting high standards of magnetic observatory practice (www.intermagnet.org).

This paper is a result of financial support in the framework of FP7, Marie Curie Actions, International Research Staff Exchange Scheme, Project title: **Complex Research of Earthquake's Forecasting Possibilities, Seismicity and Climate Change Correlations**, Acronym: **BlackSeaHazNet**, Grant Agreement Number**:** PIRSES-GA-2009-246874